\documentclass[final]{IEEEtran}
\usepackage[latin9]{inputenc}
\usepackage{bm}
\usepackage{amsmath}
\usepackage{amssymb}
\usepackage{stmaryrd}
\usepackage{graphicx}

\makeatletter
\usepackage{amsmath}
\usepackage{amssymb}

\usepackage[level=3]{wgroup_message}  
\usepackage{graphicx,psfrag,cite,subfigure}

\author{

Hao~Sun, Weiming Huang,~\IEEEmembership{Student Member,~IEEE}, Xianghao~Yu,~\IEEEmembership{Senior Member,~IEEE},
Junting~Chen,~\IEEEmembership{Member,~IEEE}
\thanks{The work was supported in part by the National Science Foundation of China (NSFC) under Grant No. 62171398, by the Basic Research Project No. HZQB-KCZYZ-2021067 of Hetao Shenzhen-HK S\&T Cooperation Zone, by NSFC Grant No. 62293482, by the Shenzhen Science and Technology Program under Grant No. JCYJ20220530143804010, No. KJZD20230923115104009, and No. KQTD20200909114730003, by Guangdong Research Projects No. 2019QN01X895, No. 2017ZT07X152, and No. 2019CX01X104, by the Shenzhen Outstanding Talents Training Fund 202002, by the Guangdong Provincial Key Laboratory of Future Networks of Intelligence (Grant No. 2022B1212010001), by the National Key R\&D Program of China with grant No. 2018YFB1800800, and by the Key Area R\&D Program of Guangdong Province with grant No. 2018B030338001.}
}


\usepackage[acronym]{glossaries}
\newcommand{\newac}{\newacronym}
\newcommand{\ac}{\gls}
\newcommand{\Ac}{\Gls}

\makeglossaries

\newac{speb}{SPEB}{square position error bound}
\newac[plural=EFIMs,firstplural=Fisher information matrices (EFIMs)]{efim}{EFIM}{Fisher information matrix}
\newac{ne}{NE}{Nash equilibrium}
\newac{mse}{MSE}{mean squared error}
\newac{toa}{TOA}{time of arrival}
\newac{tdoa}{TDOA}{time difference of arrival}
\newac{snr}{SNR}{signal-to-noise ratio}
\newac{lan}{LAN}{local area network}
\newac{psd}{PSD}{positive semidefinite}
\newac{pd}{PD}{positive definite}
\newac{wrt}{w.r.t.}{with respect to}
\newac{lhs}{L.H.S.}{left hand side}
\newac{wp1}{w.p.1}{with probability 1}
\newac{kkt}{KKT}{Karush-Kuhn-Tucker}
\newac{wlog}{w.l.o.g.}{without loss of generality}
\newac{mle}{MLE}{maximum likelihood estimation}
\newac{gps}{GPS}{global positioning system}
\newac{rssi}{RSSI}{received signal strength indicator}
\newac{mimo}{MIMO}{multiple-input multiple-output}
\newac{csi}{CSI}{channel state information}
\newac{fdd}{FDD}{frequency division duplexing}
\newac{ms}{MS}{mobile station}
\newac{bs}{BS}{base station}
\newac{d2d}{D2D}{device-to-device}
\newac{slnr}{SLNR}{signal-to-interference-leakage-and-noise-ratio}
\newac{ula}{ULA}{uniform linear antenna array}
\newac{pas}{PAS}{power angular spectrum}
\newac{mmse}{MMSE}{minimum mean square error}
\newac{zf}{ZF}{zero-forcing}
\newac{rzf}{RZF}{regularized zero-forcing}
\newac{as}{AS}{angular spread}
\newac{ue}{UE}{user equipment}
\newac{ues}{UEs}{user equipments}
\newac{iid}{i.i.d.}{independent and identically distributed} 
\newac{sinr}{SINR}{signal-to-interference-and-noise ratio}
\newac{tdd}{TDD}{time-division duplex}
\newac{rvq}{RVQ}{random vector quantization}
\newac{rhs}{R.H.S.}{right hand side}
\newac{mrc}{MRC}{maximum ratio combining}
\newac{cdf}{CDF}{cumulative distribution function}
\newac{a.s.}{a.s.}{almost surely}
\newac{los}{LOS}{line-of-sight}
\newac{jsdm}{JSDM}{joint spatial division and multiplexing}
\newac{map}{MAP}{maximum a posteriori}
\newac{klt}{KLT}{Karhunen-Lo\`eve Transform}
\newac{lbe}{LBE}{link bargaining equilibrium}
\newac{se}{SE}{Stackelberg equilibrium}
\newac{uav}{UAV}{unmanned aerial vehicle}
\newac{nlos}{NLOS}{non-line-of-sight}
\newac{pdf}{PDF}{probability density function}
\newac{em}{EM}{expectation-maximization}
\newac{knn}{KNN}{$k$-nearest neighbors}
\newac{svd}{SVD}{singular value decomposition}
\newac{nmf}{NMF}{non-negative matrix factorization}
\newac{umf}{UMF}{unimodality-constrained matrix factorization}
\newac{rmse}{RMSE}{rooted mean squared error}
\newac{olos}{OLOS}{obstructed line-of-sight}
\newac{mmw}{mmW}{millimeter wave}
\newac{ber}{BER}{bit error rate}
\newac{rss}{RSS}{received signal strength}
\newac{lp}{LP}{linear program}
\newac{ufw}{U-FW}{unimodal Frank-Wolfe}
\newac{utf}{UTF}{unimodality-constrained tensor factorization}
\newac{fw}{FW}{Frank-Wolfe}
\newac{iot}{IoT}{Internet-of-Things}
\newac{mae}{MAE}{mean absolute error}
\newac{crb}{CRB}{Cram\'er-Rao bound}
\newac{aoa}{AoA}{angle of arrival}
\newac{wcl}{WCL}{weighted centroid localization}
\newac{slf}{SLF}{spatial loss function}
\newac{btd}{BTD}{block-term tensor decomposition}
\newac{tps}{TPS}{thin plate spline}
\newac{nmse}{NMSE}{normalized mean squared error}
\newac{svt}{SVT}{singular value thresholding}
\newac{vae}{VAE}{variational autoencoder}
\newac{gan}{GAN}{generative adversarial network}
\newac{aod}{AOD}{angle of departure}
\newac{rbf}{RBF}{radial basis function}
\newac{loocv}{LOOCV}{leave one out cross validation}
\newac{lpr}{LPR}{local polynomial regression}
\newac{xl-mimo}{XL-MIMO}{extremely large multiple-input multiple-output}
\newac{6g}{6G}{sixth-generation}
\newac{cpd}{CPD}{canonical polyadic decomposition}
\newac{nnm}{NNM}{nuclear norm minimization}
\newac{mimo-ofdm}{MIMO-OFDM}{multiple-input multiple-output orthogonal frequency division multiplexing}
\newac{mimo-cdma}{MIMO-CDMA}{multiple-input multiple-output code division multiple access}
\newac{pad}{PAD}{power-angular-delay}
\newac{hmm}{HMM}{hidden Markov model}

\setkeys{Gin}{width=1.0\columnwidth}
\renewcommand{\IEEEQED}{\qed}

\makeatother

\begin{document}
\title{A Support Vector Approach in Segmented Regression for Map-Assisted
Non-Cooperative Source Localization}
\maketitle
\begin{abstract}
This paper presents a non-cooperative source localization approach
based on \ac{rss} and 2D environment map, considering both \ac{los}
and \ac{nlos} conditions. Conventional localization methods, e.g.,
\ac{wcl}, may perform bad. This paper proposes a segmented regression
approach using 2D maps to estimate source location and propagation
environment jointly. By leveraging topological information from the
2D maps, a support vector-assisted algorithm is developed to solve
the segmented regression problem, separate the \ac{los} and \ac{nlos}
measurements, and estimate the location of source. The proposed method
demonstrates a good localization performance with an improvement of
over 30\% in localization \ac{rmse} compared to the baseline methods.
\end{abstract}


\section{Introduction}

Non-cooperative localization \cite{YapLevKutCai:J23,SunChe:C21}
finds important applications in real-world scenarios. In communication
networks, accurate localization of node failures plays a crucial role
in maintaining network resilience and enhancing operational efficiency
\cite{MaLHeTSwaTow:J17}. In modern power systems, rapid and precise
localization of attacks or disturbances is essential for safeguarding
grid stability and ensuring uninterrupted service \cite{NudNabCha:J15}.
In cognitive radio networks, reliable spectrum sensing enables secondary
users to locate and access available spectrum without interfering
with primary users \cite{SheWan:J21}. In these scenarios, it is
difficult to perform localization that requires cooperation among
nodes. For example, measuring the \ac{toa} or \ac{tdoa} require
pilot sequences or preambles for an accurate estimation of the timing
of the received signal. By contrast, \ac{rss} or \ac{aoa} based
methods can detect and localize non-cooperative sources, because they
do not require the preamble or data of the signal source, but rely
on fitting the statistics of the received signal to the propagation
model.

However, \ac{rss} or \ac{aoa} based methods \cite{ZheSheLiuLiJ:J19,TomBekDin:J17}
suffer from poor localization accuracy. To begin with, these methods
require an empirical propagation model, but the model parameters may
not be accurately known by the system. In addition, as the signal
strength decreases substantially as the distance increases, a small
fluctuation in signal strength, due to multi-paths or shadowing, can
be translated to a huge difference in the ranging result. For \ac{aoa}
based methods, the accuracy of the \ac{aoa} estimation depends on
the antenna array configuration and the angular spread of the signals
due to the multi-paths. Furthermore, these methods are significantly
affected by \ac{nlos} propagation, which induces huge fluctuation
in signal strength and the spread of \ac{aoa}.

Leveraging the topological structure of radio maps can enable the
localization of a non-cooperative signal source without relying on
accurate propagation models. Here, radio maps refer to the spatial
distribution of the \ac{rss} of the signal emitted from the source
location. Thus, reversely, one can leverage the \ac{rss} patterns
and the spatial relationship with the geometric layout of the propagation
environment to infer the location of the signal source. A common scenario
is to collect the RSS of a signal source by a group of measurements
scattered at various locations. A simplest approach to estimate the
source location is to use the \ac{wcl} algorithm, which estimates
the source location as the weighted sum of the sensor locations using
the RSS as the weights \cite{MagGioKanYu:J18,MarKanGioChi:C12,WanUrrHanCab:J11}.
Another non-parametric approach is to use matrix or tensor models
for radio map representation \cite{SunChe:J22,ShrFuHong:J22,SunChe:J24},
where a radio map is first reconstructed using sparse matrix completion,
and then, the source is localized by extracting some feature vectors
from matrix factorization. Furthermore, when there is NLOS, one can
jointly reconstruct the environment and the radio map \cite{LiuChe:J23,CheChe:J24},
and by knowing the location of propagation obstacle, a substantially
better RSS-based localization performance can be achieved \cite{EsrGanGes:J21,LiuChe:J23}.

It is not surprising that exploiting 3D environment map can enhance
RSS-based localization, as the subset of RSS measurements can be identified
based on the 3D map. However, as an accurate 3D environment map is
more difficult to obtain compared to its 2D counterpart, it is important
to understand {\em how to localize a source using 2D environment maps}.
The main challenge is whether we need to jointly source location and
the heights of the obstacles for a classification of LOS and NLOS.
This paper finds that it is not necessary to reconstruct the heights
of the obstacles for map-assisted RSS-based source localization. Instead,
we develop a support vector method to exploit the information from
a 2D environment map. Specifically, we exploit the geometric property
of the propagation environment and formulate a segmented regression
problem to learn the spatial feature of the radio map. We develop
a support vector-assisted approach to solve the segmented regression
problem. We demonstrate that the proposed method offers a reliable
and effective solution for localization, maintaining high accuracy
across different number of measurements and variable shadowing conditions
and achieving a reduction in localization \ac{rmse} of over $30$\%
compared to baseline methods, making it well-suited for practical
applications requiring precise localization.

\begin{figure*}
\subfigure[]{\includegraphics[width=0.33\textwidth]{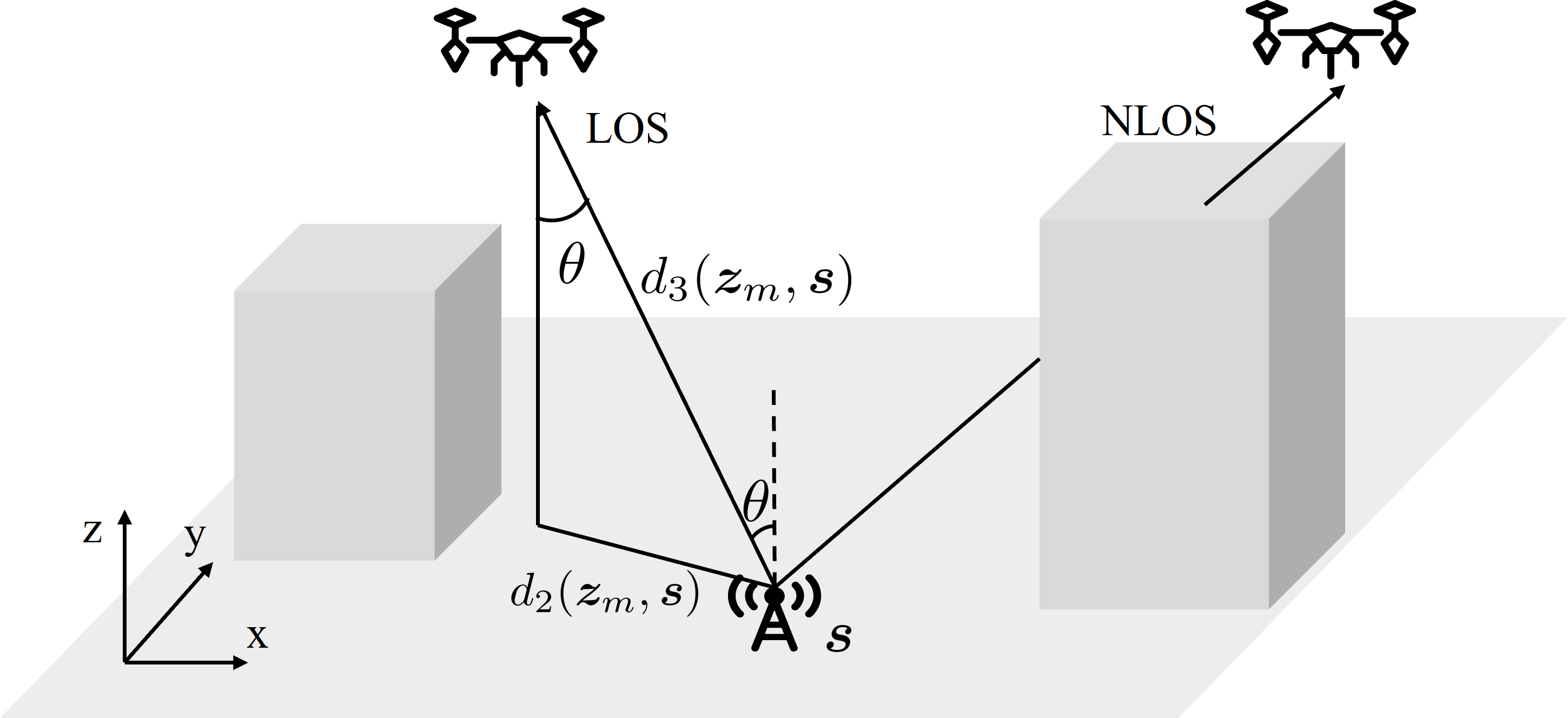}}\subfigure[]{\includegraphics[width=0.33\textwidth]{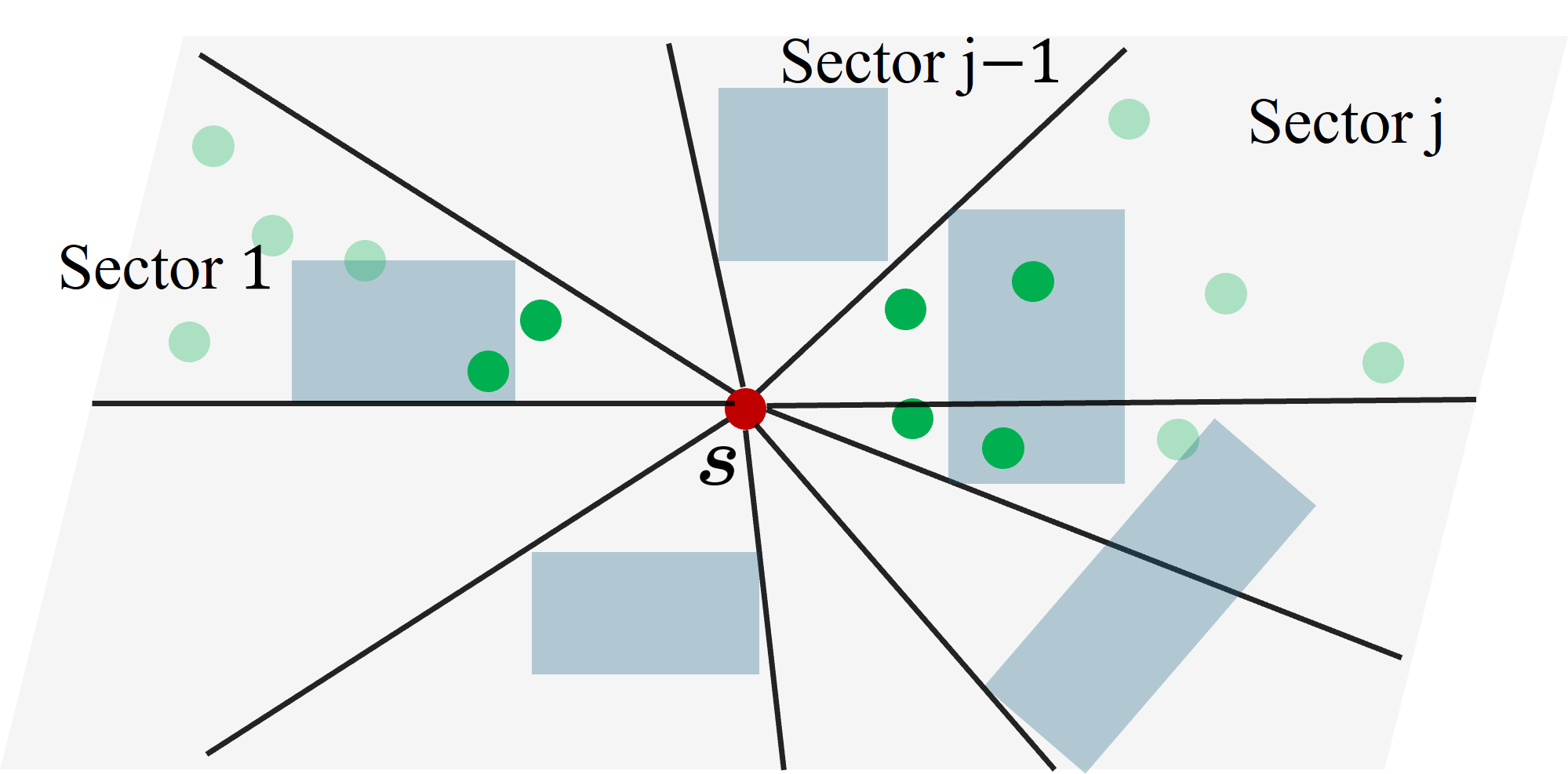}}\subfigure[]{\includegraphics[width=0.33\textwidth]{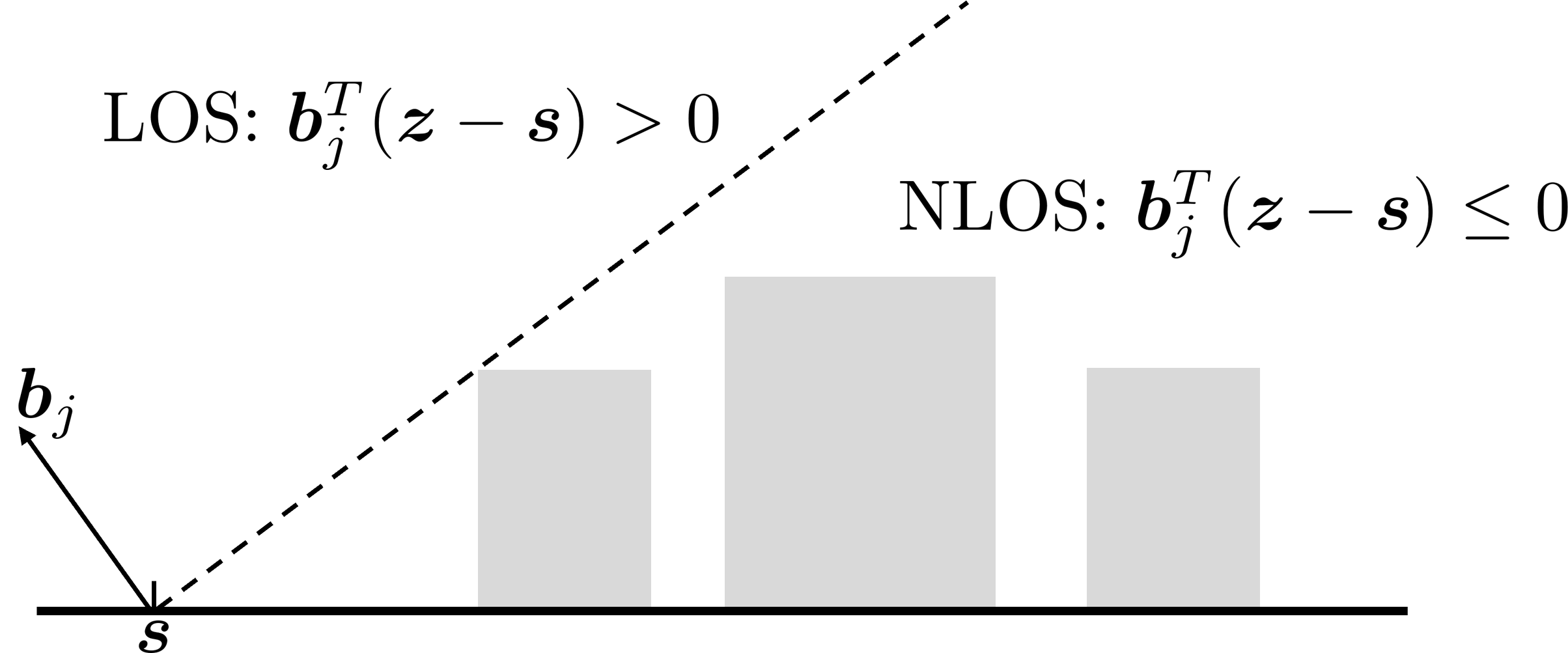}}

\caption{\label{fig:scenario}(a) A signal source is located on the ground,
and a UAV flies through the air to collect the RSS, with both LOS
and NLOS conditions present. (b) Sectors partition. Sector $j$ serves
as an example. The red circle represents the presumed source location
$\bm{s}$, the blue rectangular represents the buildings, the dark
green circle represents measurements under LOS conditions, and the
light green circle represents measurements under NLOS conditions.
(c) Support vector $\bm{b}_{j}$ and the separation between LOS and
NLOS conditions.}
\end{figure*}

\section{System Model}

Consider a blockage environment with a hidden signal source on the
ground at $\bm{s}\in\mathbb{R}^{3}$ to be localized as depicted in
Fig.~\ref{fig:scenario} (a). The locations of the propagation obstacles
are available, but the heights of the obstacles are unknown. Thus,
the blockage status for the measurement locations at some regions
is not immediately available.

Consider a set of measurements $y_{m}$ collected at different locations
$\bm{z}_{m}\in\mathbb{R}^{3}$, $m=1,2,\dots,M$, where the $m$th
measurement location can be attained by an aerial node above the ground.
In general, the measurement $y_{m}$ can be a vector if the receiver
equips an antenna array or has the capability to identify multipaths.
For the ease of elaboration, we simply consider $y_{m}\in\mathbb{R}$
measures the \ac{rss} of the signal emitted from unknown location.

\subsection{Radio Map Model}

Since there is possible blockages, denote $\mathcal{D}_{0}\subset\mathbb{R}^{6}$
as the \ac{los} propagation region for the receiver at location $\bm{z}\in\mathbb{R}^{3}$
and the signal source at $\bm{s}\in\mathbb{R}^{6}$. Likewise, $\mathcal{D}_{1}\subset\mathbb{R}^{6}$
denotes the NLOS propagation region. The radio map for each receiver
and signal source pair $(\bm{z},\bm{s})\in\mathbb{R}^{6}$ is modeled
as follows:
\begin{align}
f(\bm{z},\bm{s};\bm{\Phi}) & =\begin{cases}
f_{0}(\bm{z},\bm{s};\bm{\phi}_{0})+\epsilon_{0} & (\bm{z},\bm{s})\in\mathcal{D}_{0}\\
f_{1}(\bm{z},\bm{s};\bm{\phi}_{1})+\epsilon_{1} & (\bm{z},\bm{s})\in\mathcal{D}_{1}
\end{cases}\label{eq:radio map model}
\end{align}
where $\epsilon_{k}$, $k=0,1$ is a random variable to capture the
shadowing due to signal blockage, reflection and diffraction, etc,
and $\epsilon_{0}\sim\mathcal{N}(0,\sigma_{0}^{2})$, $\epsilon_{1}\sim\mathcal{N}(0,\sigma_{1}^{2})$,
$\bm{\Phi}=\{\mathcal{D}_{0},\mathcal{D}_{1},\bm{\phi}_{0},\bm{\phi}_{1}\}$
is a collection of the radio map parameters. The functions $f_{0}(\bm{z},\bm{s};\bm{\phi}_{0})$
and $f_{1}(\bm{z},\bm{s};\bm{\phi}_{1})$ represent the propagation
models with parameters $\bm{\phi}_{0}$ and $\bm{\phi}_{1}$, corresponding
to LOS and NLOS, respectively, and these models can be represented
by a neural network, a non-parametric model, or a parametric model
according to the application scenario to be specified later.

The aim of the model (\ref{eq:radio map model}) is to decompose the
radio map model into the component $\{\mathcal{D}_{0},\mathcal{D}_{1}$\},
which captures the LOS and NLOS patterns, and the component $\{f_{0},f_{1}\}$,
which captures the propagation law affected by unknown factors, including
power and antenna configurations. The radio map approach thus aims
at localizing the source location without explicitly recovering these
propagation factors. A general localization problem can be formulated
as follows:
\[
\underset{\bm{s},\bm{\text{\ensuremath{\Phi}}}}{\text{minimize}}\sum_{m}(y_{m}-f(\bm{z}_{m},\bm{s};\bm{\Phi}))^{2}.
\]
The problem is to find a source location $\bm{s}$ such that the corresponding
radio map $f(\bm{z}_{m},\bm{s};\bm{\Phi})$ for the source at $\bm{s}$
matches with the measurements $\{(\bm{z}_{m},y_{m})\}$.

\subsection{Segmented Propagation with Support Vectors}

It is challenging to specify the propagation regions $\mathcal{D}_{0}$
and $\mathcal{D}_{1}$, because they can appear with an arbitrary
shape. However, when the obstacle location is available, the propagation
region can be specified using support vectors that can be learned
from the measurements $\{(\bm{z}_{m},y_{m})\}$.

\subsubsection{Sectoring of Measurements}

Suppose that we are given a source location $\bm{s}$. Then, on a
top view, the area can be partitioned into $J(\bm{s})$ sectors centered
at $\bm{s}$, where these sectors approximately separate the obstacles
from each other as much as possible according to $\bm{s}$ as depicted
in Fig.~\ref{fig:scenario} (b). Then, the measurements can be clustered
to different index set $\mathbb{M}_{j}(\bm{s})$ according to $\bm{s}$
where $j$ corresponds to the sector.

In practice, sectorization is a challenging task in a complex environment.
However, in our specific application as shown in Fig.~\ref{fig:scenario}
(c) to be discussed later, there is only one building in each direction
that plays a critical role in determining the boundary of LOS and
NLOS regions, and this critical building, empirically, usually locates
nearby the source location $\bm{s}$. As a result, one can approximately
perform sectorization based on a few buildings near the source location
$\bm{s}$, although the building heights are assumed unknown. Note
that the sectorization depends on the presumed source location $\bm{s}$,
and hence, it needs to be updated with the localization algorithm
discussed later.

\subsubsection{Support Vectors and Regression Model}

The advantage of the sectoring is that for each sector $j$, the LOS
region can be determined by a linear operation with a support vector
$\bm{b}_{j}$ as depicted in Fig.~\ref{fig:scenario} (c) from a
horizontal view. Mathematically, given a presumed source location
$\bm{s}$ and defining a normal vector $\bm{b}_{j}$ of the LOS-NLOS
separating plane, the LOS region in the $j$th sector is the set of
locations $\bm{z}$ such that $\bm{b}_{j}^{\text{T}}(\bm{z}-\bm{s})>0$,
and the NLOS region in the $j$th sector is the set of locations $\bm{z}$
such that $\bm{b}_{j}^{\text{T}}(\bm{z}-\bm{s})\leq0$.

As a result, the radio map model parameter $\bm{\Phi}$ can be simplified
to $\bm{\Phi}=\{\{\bm{b}_{j}\},\bm{\phi}_{0},\bm{\phi}_{1}\}$, which
depends on the presumed source location $\bm{s}$ to be optimized.
The source localization problem becomes a joint segmented regression
problem assisted by support vectors:
\begin{align}
\underset{\bm{s},\bm{\phi}_{k},\{\bm{b}_{j}\}}{\text{minimize}} & \quad\sum_{j=1}^{J(\bm{s})}\sum_{m\in\mathbb{M}_{j}(\bm{s})}(y_{m}-\sum_{k=0}^{1}f_{k}(\bm{z}_{m},\bm{s};\bm{\phi}_{k}))^{2}u_{m}^{(k)}\label{eq:segment regression-general}\\
\text{subject to} & \quad u_{m}^{(k)}=\mathbb{I}_{k}(\bm{z}_{m},\bm{s},\bm{b}_{j}),k=0,1\nonumber 
\end{align}
where $u_{m}^{(k)}$ are auxiliary variables to classify measurements
locations $\bm{z}_{m}$ into LOS regions or NLOS regions according
to the support vectors $\bm{b}_{j}$, and
\[
\mathbb{I}_{k}(\bm{z}_{m},\bm{s},\bm{b}_{j})=\begin{cases}
\mathbb{I}(\bm{b}_{j}^{\text{T}}(\bm{z}_{m}-\bm{s})\geq0) & \text{if}\ k=0\\
\mathbb{I}(\bm{b}_{j}^{\text{T}}(\bm{z}_{m}-\bm{s})<0) & \text{if}\ k=1
\end{cases}
\]
and $\mathbb{I}(x)$ is an indicator function with $\mathbb{I}(x)=1$
if $x$ is true, and $\mathbb{I}(x)=0$, otherwise.

\section{Segmented Regression for Localization}

In this section, we consider a scenario where the source and areal
nodes all equipped with antennas for signal transmission and reception.
We propose a parametric model to represent the radio map. The parametric
model can vary with different types of antennas. Then, we propose
a segmented regression method to fit the measurements to the parametric
model to account for both LOS and NLOS conditions, which are affected
by the obstruction of buildings. Finally, the source location is estimated
by identifying the presumed location where the regression residual
is minimized.

\subsection{Parametric Model to Represent $f(\bm{z},\bm{s};\bm{\bm{\Phi}})$}

We propose a parametric model $\rho_{k}(\bm{z},\bm{s};\bm{\phi}_{k})$
to approximate the radio map model $f(\bm{z},\bm{s};\bm{\bm{\Phi}})$
in (\ref{eq:radio map model}) as follows:
\begin{align}
\rho_{k}(\bm{z},\bm{s};\bm{\phi}_{k}) & =a_{k}+b_{k}\log(d_{3}(\bm{z},\bm{s}))+c_{k}\text{log}(d_{2}(\bm{z},\bm{s}))+\epsilon_{k}\label{eq:parametric-1}
\end{align}
where $\bm{\phi}_{k}=[a_{k},b_{k},c_{k}]$.

Note that the parametric form in (\ref{eq:parametric-1}) is universal
in modeling wireless propagation channels. For example, considering
a classical empirical channel model in the linear scale
\begin{equation}
y(\bm{z})=\begin{cases}
Pg_{0}(d_{3}(\bm{z},\bm{s}))g(\theta(\bm{z},\bm{s}))+\epsilon_{0} & (\bm{z},\bm{s})\in\mathcal{D}_{0}\\
Pg_{1}(d_{3}(\bm{z},\bm{s}))g(\theta(\bm{z},\bm{s}))+\epsilon_{1} & (\bm{z},\bm{s})\in\mathcal{D}_{1}
\end{cases}\label{eq:measurement model-1}
\end{equation}
where $P$ is the transmitted power, 
\begin{equation}
g_{k}(d_{3}(\bm{z},\bm{s}))=(d_{3}(\bm{z},\bm{s}))^{-\beta_{k}}\label{eq:g_k}
\end{equation}
is the path loss and $g(\theta(\bm{z},\bm{s}))$ is the radiation
pattern of the antenna. The transmitted power $P$ corresponds to
the first term $a_{k}$ in (\ref{eq:parametric-1}). The path loss
model in the log-scale can be written as $\beta_{k}\log(d_{3}(\bm{z},\bm{s}))$
which corresponds to part of the second term $b_{k}\log(d_{3}(\bm{z},\bm{s}))$
in (\ref{eq:parametric-1}).

For the antenna radiation pattern $g(\theta(\bm{z},\bm{s}))$, we
take a vertically polarization antenna as an example. The antenna
gain is modeled as \cite{Bal:B16}:
\[
g(\theta(\bm{z},\bm{s}))=\text{cos}(\frac{\pi}{2}\text{cos}(\theta(\bm{z},\bm{s})))^{2}\approx\text{\text{sin}}^{5}(\theta)
\]
where $\text{\text{sin}}(\theta)=d_{2}(\bm{z},\bm{s})/d_{3}(\bm{z},\bm{s})$,
$d_{3}(\bm{z},\bm{s})=\|\bm{z}-\bm{s}\|_{2}$ is the distance from
the aerial node $\bm{z}$ to source location $\bm{s}$ in 3D, where
$\|\cdot\|_{2}$ denotes $l_{2}$ norm, $d_{2}(\bm{z},\bm{s})$ captures
the distance between the aerial node $\bm{z}$ and source location
$\bm{s}$ in 2D without considering the height, and $\text{log}(\text{sin}(\theta))=-\log(d_{3}(\bm{z},\bm{s}))-\text{log}(d_{2}(\bm{z},\bm{s}))$.
As a result, under the log-scale, the antenna pattern corresponds
to the third term $c_{k}\text{log}(d_{2}(\bm{z},\bm{s}))$ and part
of the second term $b_{k}\log(d_{3}(\bm{z},\bm{s}))$ in (\ref{eq:parametric-1}).

As seen from the above examples, the parametric form $\rho_{k}(\bm{z},\bm{s};\bm{\phi}_{k})$
is very general, not assuming too much information about the device,
power budget, environment, or propagation pattern.

\subsection{Segmented Regression for LOS and NLOS Measurements Separation}

In this subsection, we estimate the support vector $\bm{b}_{j}$ for
the separation of LOS and NLOS measurements in sector $j$. According
to the 2D environment map and location $\bm{s}$, the measurements
can be partitioned into totally $J(\bm{s})$ sectors. The measurements
in the same sector have a consistent blocking condition, i.e., for
the measurements satisfy $\bm{b}_{j}(\bm{z}_{m}-\bm{s})>0$, the LOS
condition exists, otherwise NLOS condition exists.

We propose to estimate $\bm{b}_{j}$ through minimizing the segmented
regression residual.

The model $\rho_{k}^{(j)}(\bm{z}_{m},\bm{s};\bm{\phi}_{k})$ at $j$th
sector part is the same as $\rho_{k}(\bm{z}_{m},\bm{s};\bm{\phi}_{k})$
in (\ref{eq:parametric-1}), since the parameters $a_{k}$, $b_{k}$,
$c_{k}$ are independent of sector $j$ and only $\bm{b}_{j}$ is
different. Then, the parametric model $\rho_{k}$ and support vector
$\bm{b}_{j}$ at sector $j$ can be estimated through solving
\begin{align}
\underset{\bm{\phi},\bm{b}_{j}}{\text{mininize}} & \quad\sum_{m\in\mathbb{M}_{j}(\bm{s})}(y_{m}-\sum_{k=0}^{1}\rho_{k}(\bm{z}_{m},\bm{s};\bm{\bm{\phi}}_{k}))^{2}u_{m}^{(k)}\label{eq:LS}\\
\text{subject to} & \quad u_{m}^{(k)}=\mathbb{I}_{k}(\bm{z}_{m},\bm{s},\bm{b}_{j}),k=0,1.\nonumber 
\end{align}

For the convenience of calculation, the matrix form of (\ref{eq:LS})
is formulated as follows:
\begin{equation}
\underset{\bm{\phi},\bm{b}_{j}}{\text{min}}\quad\|\bm{y}_{j}-\tilde{\bm{D}}^{\text{T}}\bm{\phi}\|_{2}^{2}\label{eq:matrix form LS}
\end{equation}
where $\bm{y}_{j}\in\mathbb{R}^{|\mathbb{M}_{j}(\bm{s})|\times1}$
is a vector containing all measurements $y_{m}$ for $m\in\mathbb{M}_{j}(\bm{s})$,
$\mathbb{M}_{j}(\bm{s})$ represents the set of measurement indices
in sector $j$, $|\mathbb{M}_{j}(\bm{s})|$ denotes the number of
elements in $\mathbb{M}_{j}(\bm{s})$, $\mathbb{M}_{j,i}$ denotes
the $i$th index in $\mathbb{M}_{j}(\bm{s})$, $\bm{\phi}=[\bm{\phi}_{0};\bm{\phi}_{1}]\in\mathbb{R}^{6\times1}$,
$\tilde{\bm{D}}=[\bm{x}_{\mathbb{M}_{j,1}}\circ\bm{I}_{\mathbb{M}_{j,1}};\cdots;\bm{x}_{\mathbb{M}_{j,|\mathbb{M}_{j}(\bm{s})|}}\circ\bm{I}_{\mathbb{M}_{j,|\mathbb{M}_{j}(\bm{s})|}}]\in\mathbb{R}^{6\times|\mathbb{M}_{j}(\bm{s})|}$,
$\bm{x}_{\mathbb{M}_{j,i}}=[\bm{D}_{\mathbb{M}_{j,i}};\bm{D}_{\mathbb{M}_{j,i}}]\in\mathbb{R}^{6\times1}$,
$\bm{D}_{\mathbb{M}_{j,i}}=[1,\text{log}(d_{3}(\bm{z}_{\mathbb{M}_{j,i}},\bm{s})),\text{log}(d_{2}(\bm{z}_{\mathbb{M}_{j,i}},\bm{s}))]^{\text{T}}\in\mathbb{R}^{3\times1}$,
$\bm{I}_{\mathbb{M}_{j,i}}=[u_{\mathbb{M}_{j,i}}^{(0)};u_{\mathbb{M}_{j,i}}^{(0)};u_{\mathbb{M}_{j,i}}^{(0)};u_{\mathbb{M}_{j,i}}^{(1)};u_{\mathbb{M}_{j,i}}^{(1)};u_{\mathbb{M}_{j,i}}^{(1)}]^{\text{T}}\in\mathbb{R}^{6\times1}$,
and `$\circ$' represents element-wise product.

Problem (\ref{eq:matrix form LS}) is unconstrained least-squares
problem, and it is convex \ac{wrt} $\bm{\phi}$. It can be solved
by setting the derivative to zero, and the solution is given by:
\[
\hat{\bm{\phi}}_{j}=(\tilde{\bm{D}}\tilde{\bm{D}}^{\text{T}})\tilde{\bm{D}}\bm{y}_{j}.
\]

Noted that the parameter $\bm{b}_{j}$ is unknown, we propose to estimate
$\bm{b}_{j}$ based on the residual of regression in (\ref{eq:matrix form LS}).
More specifically, under fixed $\bm{s}$, each value of $\bm{b}_{j}$
corresponding to a residual $\|\bm{y}_{j}-\tilde{\bm{D}}^{\text{T}}\hat{\bm{\phi}}_{j}\|_{2}^{2}$,
and the optimal $\bm{b}_{j}$ is obtained when the residual $\|\bm{y}_{j}-\tilde{\bm{D}}^{\text{T}}\hat{\bm{\phi}}_{j}\|_{2}^{2}$
is minimized.

\subsection{Localization via Regression Residual Minimization}

In this subsection, under each presumed $\bm{s}$, we solve $J(\bm{s})$
segmented regression problems together and calculate the residuals.
The location of the source is the one with the smallest regression
residual.

We propose to solve $J(\bm{s})$ segmented regressions altogether
as follows:
\begin{align}
\underset{\bm{\phi},\{\bm{b}_{j}\},\bm{s}}{\text{min}} & \quad\sum_{j=1}^{J(\bm{s})}\sum_{m\in\mathbb{M}_{j}(\bm{s})}(y_{m}-\sum_{k=0}^{1}\rho_{k}(\bm{z}_{m},\bm{s};\bm{\phi}_{k}))^{2}u_{m}^{(k)}\label{eq:sum form LS}\\
\text{subject to} & \quad u_{m}^{(k)}=\mathbb{I}_{k}(\bm{z}_{m},\bm{s},\bm{b}_{j}),k=0,1.\nonumber 
\end{align}
Similarly to (\ref{eq:matrix form LS}), the matrix form of (\ref{eq:sum form LS})
is formulated as follows:
\begin{equation}
\underset{\bm{\phi},\{\bm{b}_{j}\},\bm{s}}{\text{min}}\sum_{j=1}^{J(\bm{s})}\|\bm{y}_{j}-\tilde{\bm{D}}^{\text{T}}\bm{\phi}\|_{2}^{2}.\label{eq:sum matrix form LS}
\end{equation}

\begin{figure}
\subfigure[]{\includegraphics[width=0.5\columnwidth]{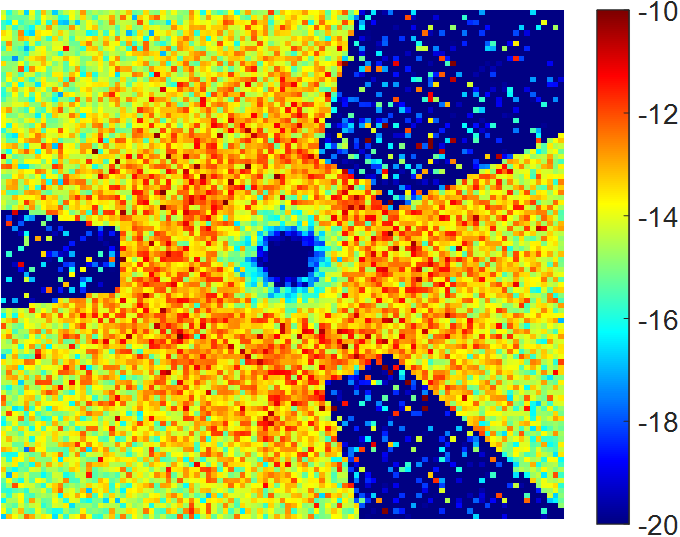}}\ \subfigure[]{\includegraphics[width=0.48\columnwidth]{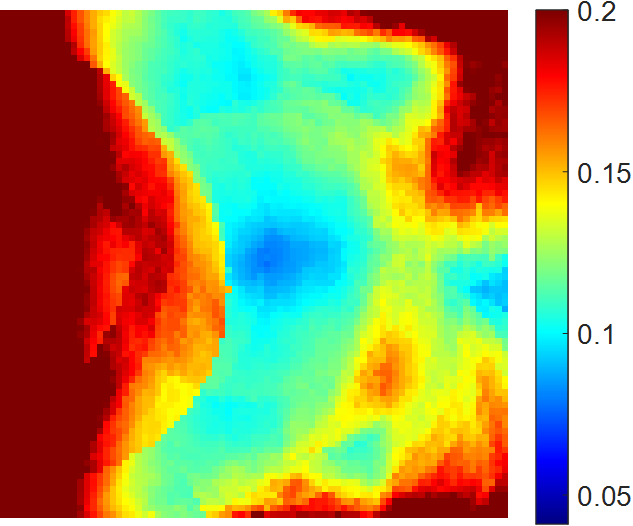}}

\caption{\label{fig:Regression-residual-at} The location of the source corresponds
to the location of the smallest regression residual. (a) The radio
map. (b) The regression residual versus location $\bm{s}$.}
\end{figure}

We propose to solve problem (\ref{eq:sum matrix form LS}) using alternating
minimization method. In problem (\ref{eq:sum matrix form LS}), when
the parameter $\{\bm{b}_{j}\}$ and $\hat{\bm{s}}$ is the ground
truth that match with the scenario, then, the solution $\hat{\bm{\phi}}$
will obtain the optimal value and the residual $\sum_{j=1}^{J(\bm{s})}\|\bm{y}_{j}-\tilde{\bm{D}}^{\text{T}}\hat{\bm{\phi}}\|_{2}^{2}$
will attain the smallest value. A visual plot of the regression residual
\ac{wrt} to the location is shown in Fig.~\ref{fig:Regression-residual-at},
 the location with smallest residual corresponds to the location of
the source. Thus, we propose to estimate $\{\hat{\bm{b}}_{j}\}$ and
$\hat{\bm{s}}$, through minimizing the residuals.

\begin{figure*}[t]
\centering{}\subfigure[]{\includegraphics[width=0.33\textwidth]{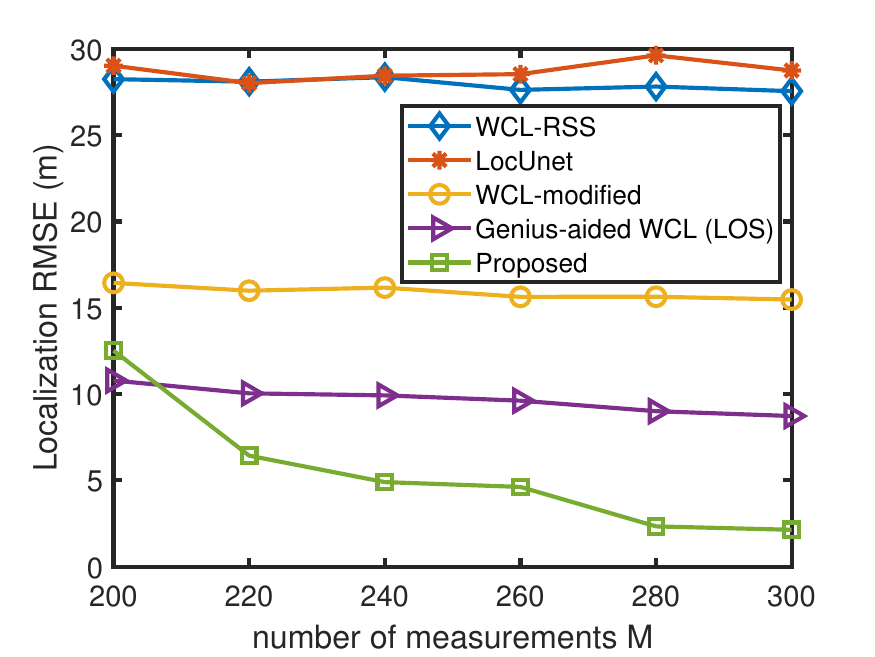}}\subfigure[]{\includegraphics[width=0.33\textwidth]{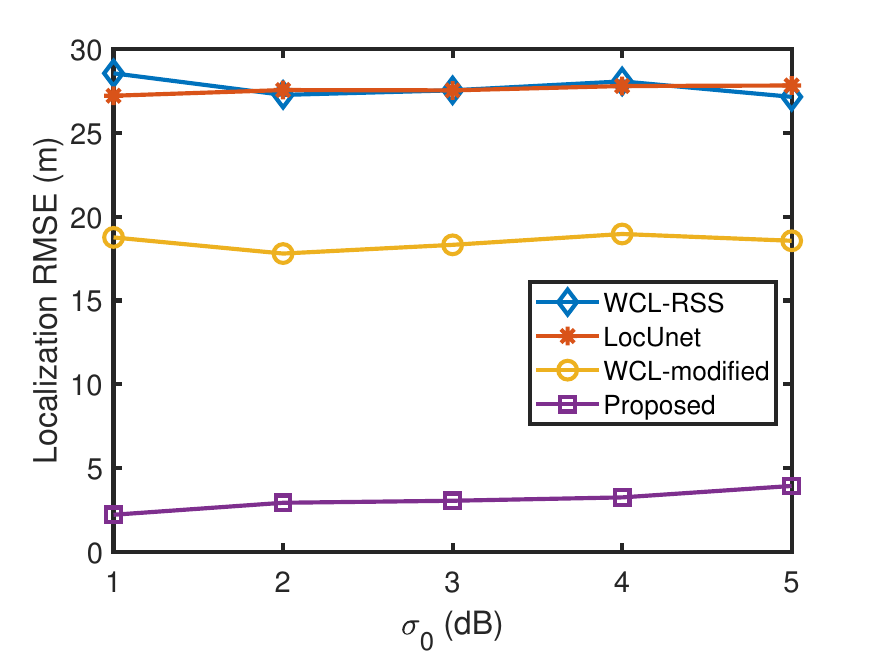}}\subfigure[]{\includegraphics[width=0.33\textwidth]{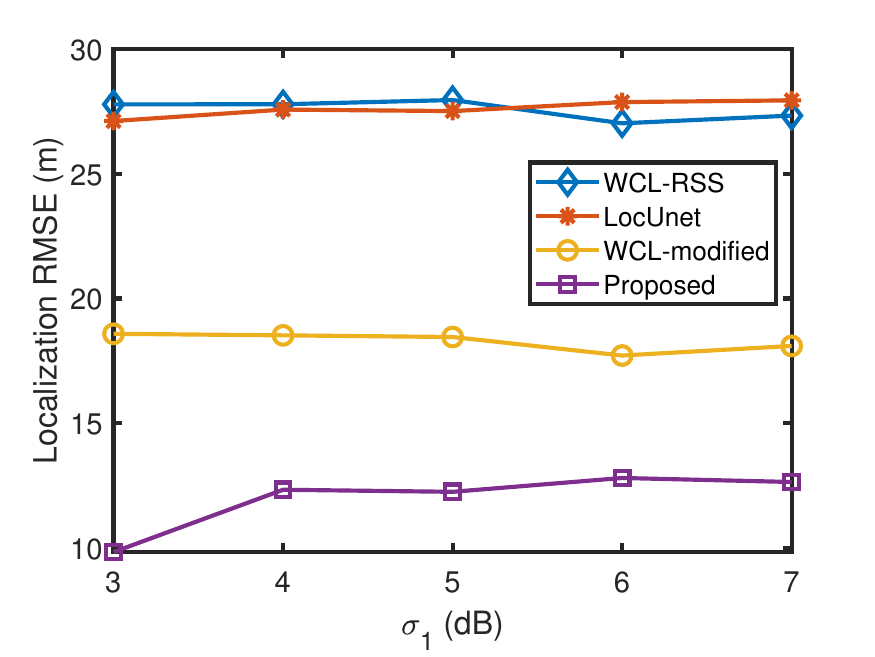}}\caption{\label{fig:Localization-RMSE-sampling-1}(a) Localization RMSE versus
number of sensors. (b) Localization RMSE versus shadowing component
$\epsilon_{0}$ under $\sigma_{0}=1-5$ dB, $\sigma_{1}=3$dB, $M=200$.
(c) Localization RMSE versus shadowing component $\epsilon_{1}$ under
$\sigma_{0}=3$dB, $\sigma_{1}=3-7$ dB, $M=200$.}
\end{figure*}
\textbf{Update of $\hat{\bm{s}}$ and $\hat{\bm{b}}_j$}: We propose
an exhaustive searching method to find the optimal $\hat{\bm{s}}$
and $\{\hat{\bm{b}}_{j}\}$ that contribute to the smallest residuals.
For each presumed $\bm{s}$ and $\bm{b}_{j}$, the residual is obtained
by $r(\bm{s},\bm{b}_{j})=\|\bm{y}_{j}-\tilde{\bm{D}}^{\text{T}}\hat{\bm{\phi}}_{j}\|_{2}^{2}$.
We can vary all the possible $N_{s}$ values of $\bm{s}$ and $N_{b}$
values of each $\bm{b}_{j}$, and construct an error tensor $\bm{\mathcal{E}}\in\mathbb{R}^{N_{s}\times N_{b}\times J(\bm{s})}$.
Then, we extract an error matrix $\bm{E}_{n_{s}}\in\mathbb{R}^{N_{b}\times J(\bm{s})}$
from $\bm{\mathcal{E}}$ where $\bm{E}_{n_{s}}=\bm{\mathcal{E}}(n_{s},:,:)$,
$n_{s}=1,\dots,N_{s}$ which represents the error under $n_{s}$th
presumed source $\bm{s}$. The error vector $\bm{E}_{n_{s}}(:,j)$
represents the error under $n_{s}$th source and $j$th sector via
varying the value of $\bm{b}_{j}$. The index of the smallest value
in $\bm{E}_{n_{s}}(:,j)$ corresponds to the optimal value $\hat{\bm{b}}_{j}$.

For each source index $n_{s}$, we calculate its corresponding error
summation as $e_{n_{s}}=\sum_{j=1}^{J(\bm{s})}\text{min}\ \bm{E}_{n_{s}}(:,j)$.
The index $n_{s}$ of the smallest $e_{n_{s}}$ corresponds to source
location $\hat{\bm{s}}$. Under the index $n_{s}$, the index $j$
of $\text{\ensuremath{\underset{j}{\text{min}}}\ }\bm{E}_{n_{s}}(:,j)$
corresponds to the optimal $\hat{\bm{b}}_{j}$.

\textbf{Update of $\bm{\phi}$}: With the estimated $\hat{\bm{s}}_{j}$
and $\{\hat{\bm{b}}_{j}\}$, solving a global coefficients $\bm{\phi}$
using all the measurements is as follows:
\begin{equation}
\underset{\bm{\phi}}{\text{min}}\|\bm{y}-\bm{\bar{D}}^{\text{T}}\bm{\phi}\|_{2}^{2}\label{eq:matrix form LS-1}
\end{equation}
where $\bm{y}=[y_{1},y_{2},\cdots,y_{M}]^{\text{T}}\in\mathbb{R}^{M\times1}$,
$\bm{\phi}=[\bm{\phi}_{0};\bm{\phi}_{1}]\in\mathbb{R}^{6\times1}$,
$\bar{\bm{D}}=[\bm{x}_{1}\circ\bar{\bm{I}}_{1};\cdots;\bm{x}_{M}\circ\bm{\bar{I}}_{M}]\in\mathbb{R}^{6\times M}$,
$\bm{x}_{m}=[\bm{D}_{m};\bm{D}_{m}]\in\mathbb{R}^{6\times1}$, $\bm{D}_{m}=[1,\text{log}(d_{3}(\bm{z}_{m},\bm{s})),\text{log}(d_{2}(\bm{z}_{m},\bm{s}))]^{\text{T}}\in\mathbb{R}^{3\times1}$,
$\bm{\bar{I}}_{m}=[u_{m}^{(0)};u_{m}^{(0)};u_{m}^{(0)};u_{m}^{(1)};u_{m}^{(1)};u_{m}^{(1)}]^{\text{T}}\in\mathbb{R}^{6\times1}$.

Problem (\ref{eq:matrix form LS-1}) is unconstrained least-squares
problem, and can be solved by setting the derivative to zero. The
solution is $\hat{\bm{\phi}}=(\bar{\bm{D}}\bm{\bar{D}}^{\text{T}})\bar{\bm{D}}\bm{y}.$

Then, the unknown parameters $\bm{\phi}$, $\{\bm{b}_{j}\}$, $\bm{s}$
in (\ref{eq:sum form LS}) are obtained.

\section{Numerical Results\label{sec:Numerical-Results}}

Consider an $L\times L$ area with $L=200$ meters and the aerial
nodes are at a height of $h=20$m. Assume the ground source $\bm{s}$
is located at coordinates $(s_{x},s_{y})$. \Ac{wlog}, we choose
$s_{x}=s_{y}=0$. To create LOS and NLOS links, assume there are three
buildings around the source. The vertices for Building 1 are $(10,40)$,
$(40,20)$, $(30,70)$, and $(60,30)$. For Building 2, the vertices
are $(80,-40)$, $(20,-80)$, $(60,-100)$, and $(100,-100)$. For
Building 3, the vertices are $(-50,20)$, $(-50,-20)$, $(-70,10)$,
and $(-70,-10)$. The aerial nodes collect measurements at $M$ locations,
chosen uniformly at random. We utilize the model (\ref{eq:parametric-1})
to generate the RSS collected by the aerial nodes. We choose $\beta_{0}=-2$
and $\beta_{1}=-7$ in (\ref{eq:g_k}) and let $P=1$W. The shadowing
component follows a Gaussian distribution, $\epsilon_{k}\sim\mathcal{N}(0,\sigma_{k}^{2})$,
with $\sigma_{0}=1$ and $\sigma_{1}=5$. We evaluate the localization
performance of the proposed method. The criterion for assessment is
the localization \ac{rmse}, calculated as $\|\hat{\bm{s}}-\bm{s}\|_{2}$,
where $\|\cdot\|_{2}$ denotes the $l_{2}$ norm. The performance
is benchmarked against four baseline methods. Baseline 1: WCL-RSS:
the source location is estimated using the formula $\hat{\bm{s}}=\sum_{m=1}^{M}w(y_{m})\bm{z}_{m}/\sum_{m=1}^{M}w(y_{m})$,
where the weight function $w(y_{m})=y_{m}$. Baseline 2: WCL-modified,
this method uses the same formula as WCL, but with a modified weight
function $w(y_{m})=y_{m}^{0.6}$. Baseline 3: Genius-aided WCL (LOS),
this method only uses the LOS measurements to perform localization.
Baseline 4: LocUnet\footnote{https://github.com/Uminan/Segmented-Regression-Localization}
\cite{YapLevKutCai:J23}, in this method, the sparse measurements
and 2D environment map are the inputs, while the location of the source
is the output.

Fig.~\ref{fig:Localization-RMSE-sampling-1} (a) illustrates the
localization RMSE for varying numbers of measurements $M$ ranging
from $200$ to $300$. The shadowing component is chosen as $\sigma_{0}=1$dB,
$\sigma_{1}=5$dB. The WCL method demonstrates poor performance due
to significant bias, which can arise when measurements are unevenly
distributed around the source. Additionally, the presence of NLOS
measurements prevents performance improvement as the number of measurements
increases. In contrast, the genius-aided WCL (LOS) method relies solely
on LOS measurements for localization, resulting in improved performance
as the number of measurements grows. The LocUnet method underperforms
due to its lack of generalization ability, as the input 2D map has
not been adequately trained. In comparison, the proposed method demonstrates
superior performance, with a substantial reduction in localization
RMSE as the number of measurements increases. This also highlights
its ability to effectively separate LOS and NLOS measurements, achieving
an improvement of over $80$\%.

Fig.~\ref{fig:Localization-RMSE-sampling-1} (b) illustrates the
relationship between localization RMSE and the shadowing component
$\epsilon_{0}$ under $\sigma_{0}=1-5$ dB and $\sigma_{1}=3$dB,
with $M=200$ measurements. The proposed method demonstrates a significant
improvement in performance, achieving over a $60$\% reduction in
RMSE compared to baseline methods.

Fig.~\ref{fig:Localization-RMSE-sampling-1} (c) illustrates the
relationship between localization RMSE and the shadowing component
$\epsilon_{1}$ under $\sigma_{1}=3-7$ dB and $\sigma_{0}=3$dB,
with $M=200$ measurements. The proposed method demonstrates a significant
improvement in performance, achieving over a $30$\% reduction in
RMSE compared to baseline methods.

\section{Conclusion}

In conclusion, this work presented a novel segmented regression approach
for non-cooperative RSS-based localization utilizing side information
from 2D environment maps. The proposed approach leverages topological
information, formulates the localization problem as a segmented regression
task, and employs a support vector-based solution to effectively estimate
the source location, even with limited measurements. The simulation
results demonstrated that the proposed method achieves over 30\% reduction
in localization RMSE compared to baseline methods under various settings.

\bibliographystyle{IEEEtran}
\bibliography{IEEEabrv,StringDefinitions,JCgroup,ChenBibCV}

\begin{thebibliography}{10}
\providecommand{\url}[1]{#1}
\csname url@samestyle\endcsname
\providecommand{\newblock}{\relax}
\providecommand{\bibinfo}[2]{#2}
\providecommand{\BIBentrySTDinterwordspacing}{\spaceskip=0pt\relax}
\providecommand{\BIBentryALTinterwordstretchfactor}{4}
\providecommand{\BIBentryALTinterwordspacing}{\spaceskip=\fontdimen2\font plus
\BIBentryALTinterwordstretchfactor\fontdimen3\font minus
  \fontdimen4\font\relax}
\providecommand{\BIBforeignlanguage}[2]{{%
\expandafter\ifx\csname l@#1\endcsname\relax
\typeout{** WARNING: IEEEtran.bst: No hyphenation pattern has been}%
\typeout{** loaded for the language `#1'. Using the pattern for}%
\typeout{** the default language instead.}%
\else
\language=\csname l@#1\endcsname
\fi
#2}}
\providecommand{\BIBdecl}{\relax}
\BIBdecl

\bibitem{YapLevKutCai:J23}
{\c{C}}.~Yapar, R.~Levie, G.~Kutyniok, and G.~Caire, ``Real-time outdoor
  localization using radio maps: A deep learning approach,'' \emph{{IEEE}
  Trans. Wireless Commun.}, vol.~22, no.~12, pp. 9703--9717, 2023.

\bibitem{SunChe:C21}
H.~Sun and J.~Chen, ``Grid optimization for matrix-based source localization
  under inhomogeneous sensor topology,'' in \emph{Proc. IEEE Int. Conf.
  Acoustics, Speech, and Signal Processing}, 2021, pp. 5110--5114.

\bibitem{MaLHeTSwaTow:J17}
L.~Ma, T.~He, A.~Swami, D.~Towsley, and K.~K. Leung, ``Network capability in
  localizing node failures via end-to-end path measurements,'' \emph{IEEE/ACM
  Trans. Netw.}, vol.~25, no.~1, pp. 434--450, 2017.

\bibitem{NudNabCha:J15}
T.~R. Nudell, S.~Nabavi, and A.~Chakrabortty, ``A real-time attack localization
  algorithm for large power system networks using graph-theoretic techniques,''
  \emph{IEEE Trans. Smart Grid}, vol.~6, no.~5, pp. 2551--2559, 2015.

\bibitem{SheWan:J21}
X.~Sheng and S.~Wang, ``Online primary user emulation attacks in cognitive
  radio networks using thompson sampling,'' \emph{{IEEE} Trans. Wireless
  Commun.}, vol.~20, no.~12, pp. 8264--8273, 2021.

\bibitem{ZheSheLiuLiJ:J19}
Y.~Zheng, M.~Sheng, J.~Liu, and J.~Li, ``Exploiting {AoA} estimation accuracy
  for indoor localization: A weighted {AoA}-based approach,'' \emph{IEEE
  Wireless Commun. Lett.}, vol.~8, no.~1, pp. 65--68, 2019.

\bibitem{TomBekDin:J17}
S.~Tomic, M.~Beko, and R.~Dinis, ``3-{D} target localization in wireless sensor
  networks using {RSS} and {AoA} measurements,'' \emph{{IEEE} Trans. Veh.
  Technol.}, vol.~66, no.~4, pp. 3197--3210, 2017.

\bibitem{MagGioKanYu:J18}
K.~Magowe, A.~Giorgetti, S.~Kandeepan, and X.~Yu, ``Accurate analysis of
  weighted centroid localization,'' \emph{IEEE Trans. on Cognitive Commun. and
  Networking}, vol.~5, no.~1, pp. 153--164, 2018.

\bibitem{MarKanGioChi:C12}
A.~Mariani, S.~Kandeepan, A.~Giorgetti, and M.~Chiani, ``Cooperative weighted
  centroid localization for cognitive radio networks,'' in \emph{Proc. Int.
  Symposium Commun. and Info. Tech.}, 2012, pp. 459--464.

\bibitem{WanUrrHanCab:J11}
J.~Wang, P.~Urriza, Y.~Han, and D.~Cabric, ``Weighted centroid localization
  algorithm: theoretical analysis and distributed implementation,''
  \emph{{IEEE} Trans. Wireless Commun.}, vol.~10, no.~10, pp. 3403--3413, 2011.

\bibitem{SunChe:J22}
H.~Sun and J.~Chen, ``Propagation map reconstruction via interpolation assisted
  matrix completion,'' \emph{{IEEE} Trans. Signal Process.}, vol.~70, pp.
  6154--6169, 2022.

\bibitem{ShrFuHong:J22}
S.~Shrestha, X.~Fu, and M.~Hong, ``Deep spectrum cartography: Completing radio
  map tensors using learned neural models,'' \emph{{IEEE} Trans. Signal
  Process.}, vol.~70, pp. 1170--1184, 2022.

\bibitem{SunChe:J24}
H.~Sun and J.~Chen, ``Integrated interpolation and block-term tensor
  decomposition for spectrum map construction,'' \emph{{IEEE} Trans. Signal
  Process.}, vol.~72, pp. 3896--3911, 2024.

\bibitem{LiuChe:J23}
W.~Liu and J.~Chen, ``{UAV}-aided radio map construction exploiting environment
  semantics,'' \emph{{IEEE} Trans. Wireless Commun.}, vol.~22, no.~9, pp.
  6341--6355, 2023.

\bibitem{CheChe:J24}
W.~Chen and J.~Chen, ``Diffraction and scattering aware radio map and
  environment reconstruction using geometry model-assisted deep learning,''
  \emph{{IEEE} Trans. Wireless Commun.}, vol.~23, no.~12, pp. 19\,804--19\,819,
  2024.

\bibitem{EsrGanGes:J21}
O.~Esrafilian, R.~Gangula, and D.~Gesbert, ``Three-dimensional-map-based
  trajectory design in {UAV}-aided wireless localization systems,'' \emph{IEEE
  Internet Things J.}, vol.~8, no.~12, pp. 9894--9904, 2021.

\bibitem{Bal:B16}
C.~A. Balanis, \emph{Antenna theory: analysis and design}.\hskip 1em plus 0.5em
  minus 0.4em\relax John wiley \& sons, 2016.

\end{thebibliography}

\end{document}